\documentclass[twocolumn,aps,prd,amssymb,eqsecnum,nofootinbib]{revtex4}
\usepackage{epsfig}
\usepackage{amsmath}
\usepackage{amssymb}

\begin{document}
\title{The conformal frame freedom in theories of gravitation}

\author{\'Eanna \'E.\ Flanagan}
\affiliation{Center for Radiophysics and Space Research, Cornell
University, Ithaca, NY 14853-5001.}

\begin{abstract}
It has frequently been claimed in the literature that the classical
physical predictions of
scalar tensor theories of gravity depend on the conformal frame in
which the theory is formulated.  We argue that
this claim is false, and that all classical physical predictions are
conformal-frame invariants.  We also respond to criticisms by Vollick
[gr-qc/0312041], in which this issue arises, of our recent analysis of
the Palatini form of 1/R gravity.
\end{abstract}

\maketitle


\section{Introduction}
\label{intro}

The observed recent acceleration of the Universe's expansion
\cite{SN,wmap} has prompted suggestions that the acceleration can be
explained by a modification of general relativity
\cite{Capo,Carroll,Chiba,Cline,Meng}.
In particular Vollick \cite{Vollick} suggested a model based on adding
a term proportional to $1/R$ to the gravitational action, where $R$ is
the Ricci scalar, and on using the Palatini form of the variational
principle.  In Ref.\ \cite{flanagan} we argued that this model is in
conflict with low-energy atomic and particle physics experiments, and
is thus ruled out.  Our analysis was criticized in a recent paper by
Vollick \cite{Vollicknew}.  In this paper we respond to
these criticisms and clarify the arguments
presented in Ref.\ \cite{flanagan}.

One of the issues that underlies and permeates the arguments of Ref.\
\cite{Vollicknew} is the nature and status of the conformal-frame
freedom in gravitation theories.  There appears to be a widespread
misconception in the literature on this issue.  Specifically, it has
often been claimed that the predictions of the theory depend on the
conformal frame in which the theory is formulated.  A secondary
purpose of this paper is to review and discuss this issue, as
background to our discussion of Vollick's criticisms.
We discuss this issue in Sec.\ \ref{sec:conformal}, and enumerate
and respond to Vollick's criticisms in Sec.\ \ref{sec:answer}.

\section{The conformal-frame freedom in theories of gravitation}
\label{sec:conformal}

\subsection{Physical theories: mathematical equivalence versus
physical equivalence}
\label{sec:equivalence}

Vollick \cite{Vollicknew} refers to the idea that two different
physical theories can be mathematically equivalent without being
physically equivalent.  There are two contexts in which
this idea makes sense.
In order to discuss those contexts, it is useful to consider
some specific examples of physical theories.

As a first example, let ${\cal T}_1$ be the standard model of particle
physics, and let
${\cal T}_2$ be the standard model with the roles of ``left-handed''
and ``right-handed'' interchanged.  Then ${\cal T}_1$ and ${\cal T}_2$
will differ due to parity violation in the weak interaction.
Are these two theories equivalent?
Clearly they are mathematically equivalent, since the states of ${\cal T}_1$
are in one-to-one correspondence with the states of ${\cal T}_2$.
They are not physically equivalent, however, from the following point
of view.  We can use specific objects exterior to the theory to define
right-handed and left-handed (for example chiral organic molecules,
whose chirality
depends on an accident of history).  With respect to this standard,
the theory ${\cal T}_1$ is correct while the theory ${\cal T}_2$
disagrees with experiments.

There is a second point of view, on the other hand, according to which
the two theories are physically equivalent.
The difference between ${\cal T}_1$ and ${\cal T}_2$
amounts to a trivial relabeling of what constitutes left-handed and
right-handed, the specification of which is an arbitrary convention.
If we were to encounter a
specification of a model of particle physics from an alien civilization, we
would not know that civilization's conventions for right-handed
and left-handed, and we would naturally say that the model was
correct if there were some choice of convention for which the theory
agreed with experiment.  More generally, in this point of view,
two theories are physically equivalent if
any arbitrary conventions used in the interpretation of the theory can
be chosen to make the two theories agree.

As a second example, consider a general nonlinear $\sigma$-model as a
classical field theory in Minkowski spacetime.  Given an
$N$-dimensional manifold $M$ with coordinates $\Phi^A = (\Phi^1,
\ldots, \Phi^N)$, a Riemannian metric $\gamma_{AB}(\Phi^C)$ and a
potential function $V$ on $M$, the action is
\begin{eqnarray}
S &=&
 \int d^4 x \bigg[
-\frac{1}{2} \gamma_{AB}(\Phi^C) \eta^{\mu\nu} \nabla_\mu \Phi^A
\nabla_\nu \Phi^B  -  V(\Phi^A)\bigg]. \nonumber \\
\label{eq:sigmamodel}
\end{eqnarray}
This theory can be expressed in different forms using different
choices of coordinates $\Phi^A$ on the manifold $M$ \footnote{
Of course if one takes the dynamical field of the theory to be the
mapping $\Phi$ from Minkowski spacetime to $M$, then a choice of coordinates
is not needed to define the action (\protect{\ref{eq:sigmamodel}});
this is the standard differential-geometric point of view.  Here, in
order to obtain a theory that ostensibly depends on the choice of
coordinates, we
define the dynamical fields of the theory to be the coordinate
representations $\Phi^A$ of the map $\Phi$.}.  The choice of
coordinates is analogous to the labeling of left-handed and
right-handed in the previous example: it
is an arbitrary convention.  Different choices of coordinates yield
theories that are mathematically equivalent.

Once again, there are two points of view on whether or not different
choices of coordinates in Eq.\ (\ref{eq:sigmamodel}) give rise to physically
equivalent theories.   In the first point of view, an observer could
in principle define a set of coordinates
$\Phi^A$ without using the properties of these fields encoded in the
action (\ref{eq:sigmamodel}), but instead using specific physical
objects in the observer's vicinity constructed out of the fields.  For
example, the observer might come across several boson stars constructed out
of different combinations of the fields, and use those stars to establish a
coordinate system.  If the observer insists on using such a definition
of the coordinates $\Phi^A$, then only one of the actions of form
(\ref{eq:sigmamodel}) will be correct, and the other ones
(corresponding to other choices of coordinates) will disagree with
observations.  This situation is analogous to using chiral organic
molecules to establish a convention for right-handedness, and
concluding that only one of the two theories ${\cal T}_1$ and ${\cal
T}_2$ discussed above is correct.

By contrast, the second point of view disallows the use of fixed conventions,
like choices of coordinates, that are exterior to the specification of
the theory.  Instead, one regards such conventions as mutable and
adjustable.  From this point of view, all choices of coordinates on
the field manifold $M$ give rise to physically equivalent theories.

It is clear from this discussion that the second point of view -- in
which arbitrary conventions are taken to be mutable and not fixed when
comparing two theories -- is the more conventional.  For example, most
physicists would immediately assert that different choices of
coordinates in the action (\ref{eq:sigmamodel}) give rise to
physically equivalent theories.  Our papers \cite{flanagan,flanagan1}
implicitly adopted this point of view, and this definition of
``physically equivalent.''  Nevertheless, the first point of view
is equally logical and consistent, albeit laden with the baggage of
a fixed, arbitrary choice of conventions.

Note that in the second point of view, a theory agrees with experiment
if there exists some choice of conventions for which the predictions
of the theory agree with observations.  Conversely, a theory can be
falsified only by showing
that for {\it all} choices of conventions, there is a disagreement with
experiment.  In particular, there is a danger of a naive analysis
that attempts to rule out a theory by showing disagreement with
experiment for only one choice of convention.
It is precisely this type of fallacy which Vollick \cite{Vollick}
claims occurs in our analysis \cite{flanagan}.
We will discuss this crucial point in detail below and explain why the
fallacy does not in fact occur.

So far, we have discussed one context in which two mathematically
equivalent theories can be regarded as physically inequivalent,
namely when there are arbitrary conventions that are fixed and that are
independent of
the specification of the theory.  A second context arises when one
gives an incomplete specification of a physical theory.  In particular,
this arises if one specifies a theory which constitutes one sector of
a larger theory, and if interactions in the larger theory
determine some conventions used in the interpretation of the smaller theory.
For example, source-free electromagnetism is mathematically equivalent
to a dual theory in which the roles of the electric and magnetic
fields have been interchanged \cite{jackson}.  This equivalence is not physical
however, since once one enlarges the theory to include couplings to
charged fields there are electric charges but no magnetic monopoles.

The conclusion we draw from this discussion is that if two theories
are mathematically equivalent, then they will always be physically
equivalent as long as (i) any arbitrary conventions arising in the
interpretation of the theory are regarded as adjustable, not fixed;
and (ii) the theory is complete and contains all the degrees of
freedom that are involved in measurements related to the theory.

\subsection{Scalar-tensor theories of gravitation: classical considerations}
\label{sec:STclassical}

We now turn to a discussion of scalar-tensor theories of gravitation
\cite{Damour92}.  In this section we will treat these theories as
classical field theories, neglecting quantum effects.  In the
following section we will discuss the extent to which the conclusions of
this section need to be modified by quantum mechanical considerations.

We start by reviewing some well-known properties of scalar-tensor theories.
The action for such theories can be written as
\begin{eqnarray}
S &=& \int d^4 x \sqrt{-g} \left[ \frac{1}{2 \kappa^2} {\cal A}(\Phi) R
  - \frac{1}{2} {\cal B}(\Phi) (\nabla \Phi)^2 - V(\Phi) \right]
\nonumber \\
\mbox{} &&+ S_{\rm m}[ e^{2 \alpha(\Phi)} g_{\mu\nu},\psi_{\rm m} ].
\label{eq:STdef}
\end{eqnarray}
Here the first term in the action depends on a metric
$g_{\mu\nu}$ and a scalar field $\Phi$; $R$ is the Ricci
scalar of $g_{\mu\nu}$ and $\kappa^2 = 8 \pi G$.  We use the sign
conventions of Ref.\ \cite{MTW}.  The second
term is the matter action $S_{\rm m}[{\hat g}_{\mu\nu},\psi_{\rm m}]$,
which is some functional of the matter fields $\psi_{\rm m}$ and of
the metric
\begin{equation}
{\hat g}_{\mu\nu} = \exp[2 \alpha(\Phi)] g_{\mu\nu}.
\label{eq:jordangeneral}
\end{equation}
This action can be taken to be the action of the standard model of
particle physics.
The theory (\ref{eq:STdef}) depends on four freely specifiable
functions of $\Phi$:
${\cal A}(\Phi)$, ${\cal B}(\Phi)$, $\alpha(\Phi)$ and the potential $V(\Phi)$.

As is well known, the form (\ref{eq:STdef}) of the theory is preserved
under a group of field redefinitions that contains two functional
degrees of freedom.  Specifically, if one defines a new
metric ${\bar g}_{\mu\nu}$ and a new scalar field ${\bar \Phi}$ via
\begin{eqnarray}
\label{eq:newvars1}
\Phi &=& f({\bar \Phi})  \\
g_{\mu\nu} &=& e^{2 \gamma({\bar \Phi})} {\bar g}_{\mu\nu}
\label{eq:newvars2}
\end{eqnarray}
for some functions $f$ and $\gamma$ with $f^\prime >0$,
then the action (\ref{eq:STdef}) can be rewritten up to a boundary term as
\begin{eqnarray}
S &=& \int d^4 x \sqrt{-{\bar g}} \left[ \frac{1}{2 \kappa^2} {\bar
    {\cal A}}({\bar \Phi}) {\bar R}
  - \frac{1}{2} {\bar {\cal B}}({\bar \Phi}) ({\bar \nabla} {\bar
    \Phi})^2 - {\bar V}({\bar \Phi}) \right]
\nonumber \\
\mbox{} &&+ S_{\rm m}[ e^{2 {\bar \alpha}({\bar \Phi})} {\bar
    g}_{\mu\nu},\psi_{\rm m} ].
\label{eq:STdef1}
\end{eqnarray}
Here the transformed functions ${\bar {\cal A}}({\bar \Phi})$,
${\bar {\cal B}}({\bar \Phi})$, ${\bar V}({\bar \Phi})$, and
${\bar \alpha}({\bar \Phi})$ are given by
\begin{subequations}
\begin{eqnarray}
\label{eq:alphatransform}
{\bar \alpha}({\bar \Phi}) &=& \alpha[f({\bar \Phi})] + \gamma({\bar
  \Phi}), \\
\mbox{} {\bar V}({\bar \Phi}) &=& e^{4 \gamma({\bar \Phi})} V[ f({\bar
  \Phi})],  \\
{\bar {\cal A}}({\bar \Phi}) &=& e^{2 \gamma({\bar \Phi})} {\cal
  A}[f({\bar \Phi})],
 \\
{\bar {\cal B}}({\bar \Phi}) &=& e^{2 \gamma({\bar \Phi})} \bigg\{
  f'({\bar \Phi})^2 {\cal  B}[f({\bar \Phi})]
- \frac{6}{\kappa^2} f'({\bar \Phi}) \gamma'({\bar \Phi})
  {\cal A}'[f({\bar \Phi})]
\nonumber \\ \mbox{} &&
- \frac{6}{\kappa^2} \gamma'({\bar \Phi})^2
  {\cal A}[f({\bar \Phi})] \bigg\}.
\label{eq:transformed}
\end{eqnarray}
\end{subequations}

For many theories the transformation group (\ref{eq:newvars1}) --
(\ref{eq:newvars2}) can
be used to obtain canonical representations of the theory,
characterized by two free functions instead of four.  Some of these
representations (or choices of ``conformal frame'') are

\begin{itemize}
\item The {\it Jordan frame}, which is characterized by
$\alpha =0$, ${\cal B} =1$.  The free functions in this frame are
${\cal A}(\Phi)$ and $V(\Phi)$.  Freely falling objects built from the
matter fields $\psi_{\rm m}$ follow geodesics of the Jordan-frame metric.

\item The {\it Einstein frame}, which is characterized by
${\cal A} =1$, ${\cal B} =1$.  The free functions in this frame are
$\alpha(\Phi)$ and $V(\Phi)$.

\item A frame which does not have a standard name that  is characterized by
${\cal B} = 0$, ${\cal A}(\Phi) = \Phi$.  The free functions in this
frame are $\alpha(\Phi)$ and $V(\Phi)$.


\end{itemize}

Suppose now that we are given an arbitrary scalar-tensor theory, specified by
the functions ${\cal A}(\Phi)$, ${\cal B}(\Phi)$, $V(\Phi)$ and
$\alpha(\Phi)$.  Then there is no guarantee that the theory has a well
defined classical dynamics, i.e. a well-posed initial value formulation
\cite{Wald}.
For example, in the theory ${\cal B}=1$, $V = \alpha =0$, ${\cal A} =
1 - \kappa^2 \Phi^2/6$ of a conformally coupled scalar field, there is
a class of solutions of the equations of motion with no matter given by
$\Phi = \sqrt{6}/\kappa$, $g_{\mu\nu} = {\rm any}$ metric with $R=0$.  Therefore
for this theory the initial value formulation is ill-posed; local
uniqueness of solutions fails.
However, it is often possible to obtain a sensible
classical theory by
passing to a subspace of the phase space defined by
restricting the allowed values of $\Phi$ in an initial data set to an
open interval of the form $(\Phi_1,\Phi_2)$ \footnote{Here we do
  not exclude $\Phi_1 = - \infty$ or $\Phi_2 = \infty$.
In the cases we consider below the condition $\Phi_1 < \Phi < \Phi_2$
will be preserved under dynamical evolution.}.  We will
restrict attention to
theories obtained by passing to subspaces of this form which are in
one-to-one correspondence with the entire phase space of an
Einstein-frame description
\footnote{This assumption is equivalent to the conditions
${\cal A} > 0$ and ${\cal F} > 0$ on the
interval $(\Phi_1,\Phi_2)$,
where
$
{\cal F} = {\cal B}/{\cal A} + 3 ( {\cal A}^\prime)^2 / (2
  \kappa^2 {\cal A}^2),
$
and
$$
\int_{{\tilde \Phi}_1}^{{\tilde \Phi}_2} \, \sqrt{{\cal
    F}(\Phi)} d \Phi \, \to\infty
$$
as ${\tilde \Phi}_1 \to \Phi_1$ from above and as ${\tilde \Phi}_2 \to
\Phi_2$ from below.}.
This assumption implies that all the theories we consider will
automatically have well-posed initial value formulations, since
general theorems \cite{Wald} guarantee this for Einstein-frame theories.
It also implies that any two theories related by transformations of the
form (\ref{eq:newvars1}) -- (\ref{eq:newvars2}) are mathematically
equivalent, so that it is natural to call such theories different
conformal-frame representations of the same theory.
Equivalent assumptions are more or less implicit in many discussions of scalar
tensor theories in the literature.

We now discuss whether or not different conformal-frame
representations of a theory are physically equivalent.
One fairly trivial context in which an apparent physical inequivalence
arises is when the theory is incompletely specified, when one
specifies only the first term in the action
(\ref{eq:STdef}) and not the second (matter) term.
This arises if one adopts a convention in which it is implicit
that the metric which is being used is the Jordan-frame metric, or
equivalently in which $\alpha(\Phi) =0$.  In this context there is no
conformal-frame freedom, since it is clearly inconsistent to perform a
conformal transformation on the first term, thereby transforming
the functions ${\cal A}(\Phi)$, ${\cal B}(\Phi)$ and $V(\Phi)$,
without keeping track of the transformation of the function
$\alpha(\Phi)$ in the second term.  Thus, while one can express the action of
the theory [the first term in Eq.\ (\ref{eq:STdef})] in different
conformal frames, only one of those frames is physically correct,
as pointed out by Brans \cite{Brans} \footnote{Some of the confusion in
the literature discussed below apparently arises from a
misinterpretation of the context of Brans's argument.}.

Putting aside this special case, turn now to
the more complete framework where the entire action
(\ref{eq:STdef}) is specified, including the four functions ${\cal
A}(\Phi)$, ${\cal B}(\Phi)$, $V(\Phi)$ and $\alpha(\Phi)$.
Are different conformal-frame representations of the theory physically
equivalent?  The action (\ref{eq:STdef}) is complete and contains all the
relevant degrees of freedom.  Also the representations are
mathematically equivalent because of our assumption
discussed above.
It therefore follows from the
general discussion of Sec. \ref{sec:equivalence} above that
the different conformal-frame representations {\it are} physically equivalent,
as long as one does not treat as fixed any convention that arises in the
interpretation of the theory.  The relevant convention in this case
is the interpretation of the meaning of the metric that appears in the
theory.  For example, suppose one measures the distance to the moon
using lunar laser ranging and using an atomic clock to determine the light
travel time.  In general relativity, the computation
corresponding to this measurement is straightforward: the atomic clock
measures proper time associated with the metric.  In a general
scalar-tensor theory and in a general conformal frame, however, the
time on the clock will not be the proper time associated with the metric.
Nevertheless one can still compute from the action (\ref{eq:STdef})
the time as measured by the clock,
and the same result will be obtained in all conformal frames.  Namely,
it will be the proper time associated with the metric
(\ref{eq:jordangeneral}), which from Eqs.\ (\ref{eq:newvars2}) and
(\ref{eq:alphatransform}) is a conformal-frame invariant.
Different conformal frames will be physically inequivalent only if one
insists on interpreting the metric in the theory as the metric which
is measured by atomic clocks.

However, it has repeatedly been claimed in the literature that
only one choice of conformal frame is correct or ``physical'', and
that as a consequence different conformal-frame representations of a
theory are {\it not}
physically equivalent
\cite{Vollicknew,magnano0,sokolowski00,Sok1,Ferraris,magnano,sokolowski.gr14,faraoni,Bellucci,Gunzig,Capo1}
\footnote{A source of confusion is that some authors use the phrase
``physical frame'' as a synonym for Jordan frame, whereas others
\protect{\cite{Vollicknew,magnano0,sokolowski00,Sok1,Ferraris,magnano,sokolowski.gr14,faraoni,Bellucci,Gunzig,Capo1}}
use it in the sense discussed here.}.
Several different criteria have been suggested to determine the
``correct'' or ``physical'' conformal frame, including local positivity of
energy \cite{sokolowski00,magnano,Gunzig} and the existence of a
stable ground state
\cite{sokolowski.gr14} \footnote{The first of these criteria, the
local positivity of $G_{\mu\nu} u^\mu u^\nu$ for timelike vectors
$u^\mu$, does pick out a unique conformal frame
\protect{\cite{sokolowski00}}.
The second criterion does not, since the existence and stability of a
suitable ground state is a conformal-frame invariant.}.
Such efforts to determine the ``correct'' choice of conformal frame
are misguided, at least in the realm of classical physics.
They are
analogous to attempting to determine the
``correct'' choice of radial coordinate in the Schwarzschild
spacetime.  In that context, there is of course no
correct radial coordinate, since all physical observables
are coordinate-invariants.  In a similar way,
all observable quantities in scalar-tensor theories are
conformal-frame invariants.  For example, the measured 4-momentum of
an isolated object is given by the Bondi 4-momentum of the
frame-invariant metric (\ref{eq:jordangeneral}).

We have given a general, abstract argument in favor of the physical
equivalence of different conformal frames.  That abstract argument is
supported by several explicit computations in the literature, where
observables have been computed in different conformal frames, and
where consistent results have been obtained.  For example, one of the
key observables of extended inflation models is the spectrum of scalar
perturbations.  This spectrum can be computed in
the Einstein frame or in the Jordan frame; the results are identical
\cite{extendedinflation,Kaiser}.
Another example is the careful computation by Armend\'ariz-Pic\'on
of the observed ratio between the frequency of a quasar absorption line
and the corresponding atomic transition frequency as measured in the
laboratory \cite{AP}.  That computation is carried out in a
theoretical context (more
general than the context considered here) which allows time variations of the
fine structure constant.  Armend\'ariz-Pic\'on computes the
ratio in both the Einstein and Jordan frames, and obtains identical
results.

Vollick \cite{Vollicknew} claims that different physical predictions
are obtained in the Einstein and Jordan frames, and cites as evidence
several papers \cite{Bellucci,magnano,faraoni,Gunzig} that compute
such differing predictions.  The first of these papers \cite{Bellucci} claims
that that in the Einstein-frame version of Brans-Dicke theory,
scalar gravitational waves produce
longitudinal accelerations in gravitational wave detectors,
and that no such longitudinal accelerations are present in the
Jordan-frame version of the theory.  In fact the prediction of both
frames is that there is no measurable longitudinal acceleration \footnote{The
error of Ref.\ \protect{\cite{Bellucci}} was to assume that the measured
acceleration in gravitational-wave detectors corresponds to the
relative acceleration of two freely falling particles computed in the
Einstein-frame metric, when in fact it is their relative acceleration
computed in the metric (\ref{eq:jordangeneral}) which is what is
measured.}.  The remaining
papers \cite{magnano,faraoni,Gunzig} show that the weak energy
condition is violated in one frame but not in the other.
While this is true, there is no physical observable whose predicted
value in all conformal frames is the sign of $G_{\mu\nu} u^\nu u^\mu$
for timelike vectors $u^\mu$.  Thus there is no measurable
inconsistency.

Another aspect of this subject is the
equivalence between the choice of a conformal frame and the choice of
physical units, first clearly enunciated by Dicke
\cite{Dicke}.  When we change from one system of
units to another, the ratio between the original unit of length and
the new unit of length is normally a constant, independent of space
and time.   However, if the operational procedure used to define the
unit of length is changed, then the ratio between the old and new
units can vary with space and time
\footnote{This is related to the
fact that a statement like ``Newton's constant of gravitation is
changing with time'' is meaningless, while the statement ``Newton's
constant of gravitation in SI units is changing with time'' does make
sense.  See Duff \protect{\cite{Duff}} for a lucid discussion of this
issue.}.  For example, suppose we define units of length and time by taking
the speed of light to be unity and by taking the unit of time to
be determined by some atomic transition frequency (as in the current SI
definition of the second).
Measurements of the geometry of spacetime in these units yield the
Jordan-frame metric.  However, we can instead define a system of units
as follows.  Suppose that we have a nonspinning black hole.  We can in
principle
take this to be a ``standard'' black hole (like the original
platinum-iridium standard meter), and create other nonspinning
black holes of the same size.  Using these black holes we can
operationally define a unit of time to be the inverse of the frequency
of their fundamental quasinormal mode of vibration.  If we define
the speed of light to be unity, and measure the geometry of spacetime
in these units, the result is the Einstein-frame metric.
Thus, the choice of conformal frame is no more than a choice of
physical units, just a human convention.

\subsection{Scalar-tensor theories of gravitation: quantum mechanical
considerations}
\label{sec:STquantum}

In classical physics, we have argued that scalar tensor theories can
be formulated in any conformal frame without affecting the physical
predictions of the theory.  In quantum physics the situation is not as
clear cut because (i) theories which are classically equivalent can be
quantum mechanically inequivalent, and (ii) there are the usual
difficulties in describing quantum gravitational degrees of freedom.
The status of the conformal-frame freedom depends to some extent on
the level of generality one aspires to, i.e., on the theoretical
framework or context in which one is working.

One possible context is a semiclassical approximation in which the
matter fields $\psi_{\rm m}$ in the action (\ref{eq:STdef}) are
treated quantum mechanically, but the fields $g_{\mu\nu}$ and $\Phi$
are treated classically.  Here the arguments of the last section
continue to apply, and identical predictions are obtained from all
conformal frames.  However, the arguments in Ref.\ \cite{flanagan}
used a quantum mechanical treatment of the field $\Phi$, so this
context is not sufficiently general for our purposes.

A second context is a semiclassical approximation in which the matter
fields $\psi_{\rm m}$ and the scalar field $\Phi$ are treated quantum
mechanically, but the metric $g_{\mu\nu}$ is treated classically.
Here it is well-known that quantization does not commute
with changing conformal frames \cite{sokolowski00,Fujii}, so that by
starting in different conformal frames one obtains theories that are
mathematically and physically inequivalent.
On the other hand, one would expect to find such conformal-frame
dependence from an approximation that freezes the quantum fluctuations
in a conformal-frame-dependent combination of the scalar and tensor degrees of
freedom.  There is no evidence in this context that the frame dependence
seen is not an artifact of the approximation being used, or that
frame invariance is not present at a more fundamental
level in a full quantum theory.

A third context is full quantum gravity.  Here
it is certainly conceivable that
applying a quantization procedure to different conformal-frame
representations of a theory will yield inequivalent quantum
theories\footnote{See for example the analysis of Ref.\
\protect{\cite{Ashtekar}} where the two quantum theories obtained are not
obviously equivalent, although the black hole entropies computed
in the two different frames agree.}\footnote{In this paper we restrict
attention to theories for which there exist some conformal frame in which
${\cal A} = {\cal B} =1$, as discussed in footnote 3 above.  However,
one could instead consider the class of scalar-tensor theories which
are exactly equivalent to general relativity at the classical level.
For these theories
$\alpha(\Phi)=\lambda(\Phi)$, $V(\Phi) = 0$, ${\cal A}(\Phi) = \exp[2
  \lambda(\Phi)]$, and ${\cal B}(\Phi)=- 6 \exp[2 \lambda(\Phi)]
\lambda'(\Phi)^2/\kappa^2$ for some function $\lambda(\Phi)$.
In Ref.\ \protect{\cite{Woodard}} it
is shown that these theories are quantum mechanically equivalent
to general relativity.}.  In fact, computations in two
dimensional dilaton gravity theories indicate that such
inequivalence does occur \cite{odintsov,wittenbh,review}.
However, the status of the conformal frame freedom in full quantum
gravity is actually irrelevant for our purposes, since the arguments
presented in Ref.\ \cite{flanagan} do not rely on this status.

A fourth context is the theoretical framework of effective field
theory, where all the degrees of freedom are treated quantum
mechanically but only low energy processes can be computed accurately.
This is the context that was implicitly assumed in Ref.\
\cite{flanagan}, and is the most appropriate and useful context for
the arguments presented there.  It can be applied to general
relativity \cite{Burgess,DonoghueRev} and to scalar-tensor theories in
the regime of small perturbations off Minkowski spacetime.
A key result in this context is the equivalence theorem, which says that the
scattering matrix is invariant under nonlinear local field redefinitions
\cite{et}.  It follows from this theorem that tree-level particle scattering
cross sections can be computed in any conformal frame, and identical
results will be obtained.  As we discuss in the next section, this
result is sufficient for the arguments of Ref.\ \cite{flanagan}.

\section{Palatini form of 1/R gravity}
\label{sec:answer}

In this section we discuss the specific criticisms of Ref.\
\cite{Vollicknew}.  We start by briefly reviewing the argument of
Ref.\ \cite{flanagan}.

\subsection{Review of derivation}

In the theory of gravity suggested in Ref.\ \cite{Vollick}, the
dynamical variables are a metric ${\bar g}_{\mu\nu}$, a symmetric
connection ${\hat \nabla}_\mu$, and the matter fields $\psi_{\rm m}$.
The action is
\begin{equation}
S[{\bar g}_{\mu\nu},{\hat \nabla}_\mu,\psi_{\rm m}] = \frac{1}{2
  \kappa^2} \int d^4 x \sqrt{- {\bar g}} f({\hat R}) +
S_{\rm m}[{\bar g}_{\mu\nu},\psi_{\rm m}],
\label{eq:action4}
\end{equation}
where ${\hat R}$ is the Ricci scalar of the connection ${\hat
\nabla}_\mu$ and $f({\hat R}) = {\hat R} - \mu^4/{\hat R}$.
In the variational principle the metric and connection are treated as
independent variables according to the Palatini prescription.
In Ref. \cite{flanagan} we derived another description of the theory
by (i) making field redefinitions; (ii) setting to zero some degrees
of freedom which vanish classically on-shell; (iii) adding a new
scalar field which vanishes classically on-shell; and (iv) discarding
a boundary
term.  The last three steps are valid classically, and also quantum
mechanically if one is only interested in computing tree-level
scattering cross sections as we do below.  The resulting action
depends on a metric $g_{\mu\nu}$ and a scalar field $\Phi$ and is
given by\footnote{Actually there are two sectors of the theory
(\protect{\ref{eq:action4}}), each with its own Einstein-frame
description.  The first sector is defined by the
requirement that the sign of the right hand side of Eq.\ (20) of
Ref.\ \protect{\cite{flanagan}} be positive, and it is described by
the action (\protect{\ref{eq:action9}}).  In the second sector
the right hand side of Eq.\ (20) of Ref.\ \protect{\cite{flanagan}} is
negative, and it is described by the action
(\protect{\ref{eq:action9}}) with the sign of the potential flipped.
The derivation of Eq.\ (\protect{\ref{eq:action11}}) below and the
subsequent arguments
are valid in both sectors.}
\begin{eqnarray}
{\tilde S}[g_{\mu\nu},\Phi,\psi_{\rm m}] &=&
\int d^4 x
\sqrt{- g} \bigg[ \frac{R}{2 \kappa^2}  - V(\Phi)
\bigg] \nonumber \\
&& + S_{\rm m}[e^{2 \alpha(\Phi)} g_{\mu\nu},\psi_{\rm m}],
\label{eq:action9}
\end{eqnarray}
where the potential $V(\Phi)$ and coupling function $\alpha(\Phi)$ are
given by Eqs.\ (9) and (11) of Ref.\ \cite{flanagan}.

A key point about the action (\ref{eq:action9}) is that the scalar field
is not an independent dynamical variable, but can be eliminated from
the action.  One then obtains a theory consisting of general
relativity coupled to a modified matter action.
Taking the matter action to be the Dirac action for free electrons,
integrating out the scalar field and rescaling the metric $g_{\mu\nu}
\to (4/3) g_{\mu\nu}$ gives
\begin{eqnarray}
&& {\tilde S}[g_{\mu\nu},\Psi_{\rm e}] =
\int d^4 x
\sqrt{- g} \bigg[ \frac{R}{2 {\tilde \kappa}^2}  - \Lambda
+ i {\bar \Psi}_{\rm e} \gamma^\mu \nabla_\mu \Psi_{\rm e}
\nonumber \\
&&
- m_{\rm e} {\bar \Psi}_{\rm e} \Psi_{\rm e} - \frac{3 \sqrt{3}}{16
  m_*^4} (i {\bar \Psi}_{\rm e} \gamma^\mu \nabla_\mu  \Psi_{\rm e})^2
- \frac{1}{\sqrt{3}} \frac{m_{\rm e}^2}{m_*^4} ( {\bar
 \Psi}_{\rm e} \Psi_{\rm e})^2
\nonumber \\
&&
+ \sqrt{\frac{3}{4}} \frac{m_{\rm e}} {m_*^4} (i {\bar \Psi}_{\rm e} \gamma^\mu
\nabla_\mu \Psi_{\rm e}) ( {\bar
 \Psi}_{\rm e} \Psi_{\rm e}) + \ldots
\bigg],
\label{eq:action11}
\end{eqnarray}
where ${\tilde \kappa} = \sqrt{4/3} \kappa$, $m_* = \sqrt{\mu /
\kappa}$ and $\Lambda = \mu^2 / (\sqrt{3} \kappa^2)$ is the
induced cosmological constant. The last three terms in Eq.\
(\ref{eq:action11}) are characterized by the energy scale $m_*$ which is
of order $10^{-3}$ eV.  Since this energy scale is so small, one expects the
action (\ref{eq:action11}) to be in conflict with atomic physics
and particle physics experiments, for example electron-electron scattering.
This is discussed in more detail in Sec.\ \ref{sec:qft} below.

\subsection{Objections to derivation}

Vollick \cite{Vollicknew} gives several objections to the above
derivation:

\subsubsection{``Failure to properly identify the physical frame''}

As we have argued in Secs. \ref{sec:STclassical} and
\ref{sec:STquantum}, we disagree that one needs to identify a
preferred ``physical'' conformal frame in order to define the
theory\footnote{We reiterate that here and in Ref.\
\protect{\cite{Vollicknew}} the phrase ``physical frame'' is not used
as a synonym for Jordan frame.  It is certainly true that a
specification of which frame is the Jordan frame [which is equivalent
to specifying the coupling function $\alpha(\Phi)$] is necessary to
define the theory.}.

\subsubsection{``Failure to ... add
the minimally coupled matter Lagrangian in [the physical] frame''}

The choice of frame in which to add the minimally coupled
matter Lagrangian is a part of the specification of the theory of
gravity, as discussed in Sec.\ \ref{sec:STclassical}.  That
specification was given in Refs.\ \cite{Carroll,Vollick} via the
action (\ref{eq:action4}).  Ref.\ \cite{flanagan} did not make any
choices in this regard; it simply analyzed the given theory of gravity.

\subsubsection{``The physical inequivalence of the Jordan and Einstein
frames ... has been shown by many authors''}

This issue was discussed in Secs.\ \ref{sec:STclassical} and
\ref{sec:STquantum}.   One additional source of confusion is the
following.  Vollick \cite{Vollicknew} gives an example of
two conformally related metrics and notes that they are not physically
equivalent.  However, in order to determine whether two different
situations are physically equivalent, one needs to specify not only the metric,
but also the scalar field and the four functions that define the theory of
gravity.  For example, consider the following three different
situations:
\begin{itemize}
\item[a.] The metric and scalar field are $g_{\mu\nu}$ and $\Phi$, and the
theory is defined by the functions ${\cal A}(\Phi)$, ${\cal B}(\Phi)$,
$V(\Phi)$ and $\alpha(\Phi)$.

\item[b.] The metric and scalar field are ${\bar g}_{\mu\nu}$ and ${\bar
\Phi}$, related to $g_{\mu\nu}$ and $\Phi$ by Eqs.\ (\ref{eq:newvars1}) --
(\ref{eq:newvars2}), and the theory is defined by the same functions
  ${\cal A}$, ${\cal B}$, $V$ and $\alpha$ as before.

\item[c.] The metric and scalar field are ${\bar g}_{\mu\nu}$ and ${\bar
\Phi}$, and the theory is defined by the functions
  ${\bar {\cal A}}$, ${\bar {\cal B}}$, ${\bar V}$ and ${\bar \alpha}$
given by Eqs.\ (\ref{eq:transformed}).

\end{itemize}
Here a.\ and b.\ are not physically equivalent, as noted by Vollick.
However a.\ and c.\ are physically equivalent, and it is this
which underlies the equivalence between the Einstein
and Jordan frames.

\bigskip

\subsubsection{``One cannot add a minimally coupled Lagrangian in one frame
and then transform to another frame ... all predictions of the theory
must be calculated in [the physical] frame''}

We disagree with these assertions.  As we have argued in Secs.\
\ref{sec:STclassical} and \ref{sec:STquantum}, all physical
observables are conformal-frame-invariants, and so can be computed in
any convenient frame.  This is true classically, and also quantum
mechanically for tree-level scattering amplitudes.
These assertions would only be true if one insisted on
associating particular operational meanings to the dynamical
fields appearing in the Lagrangian.  In the point of view
we adopt here [the second of the two points of view discussed in Sec.\
\ref{sec:equivalence}], we remain agnostic as to the physical
interpretation of the dynamical fields \cite{AP}, and as a consequence
we can perform computations in any conformal frame.

However, the spirit of this objection can be translated into our point
of view.  It can be summarized as ``In the
action (\ref{eq:action11}), how do we know that the fields
$g_{\mu\nu}$ and $\Psi_{\rm e}$ correspond to the graviton and electron
that we actually measure?  How do we know that there is not some
nonlinear local field redefinition which defines ``graviton'' and
``electron'' fields as nonlinear functions of $g_{\mu\nu}$
and $\Psi_{\rm e}$ in such a way that the action becomes the conventional
standard model action when expressed in terms of these new fields?''
This is a valid concern (c.f. the discussion near the end of Sec.\
\ref{sec:equivalence} above) which we now discuss.
Consider the specific case of the scattering of two non-relativistic
electrons.  The key point here is the equivalence theorem discussed in
Sec. \ref{sec:STquantum} above \cite{et}.  This theorem guarantees that the
electron-electron scattering cross section is invariant under
nonlinear local field redefinitions.  Thus, we are free to use the form
(\ref{eq:action11}) of the action to compute that cross section.

\subsection{Confrontation with experiment}
\label{sec:qft}

In this section we discuss in more detail the predictions of the
action (\ref{eq:action11}) and the extent to which they disagree with
experiments, and we correct an error in the corresponding discussion
in Ref.\ \cite{flanagan}.  We use units in which $\hbar=1$ but $c \ne
1$.

We focus attention on the interactions of non-relativistic electrons.
For this purpose it is a good approximation to neglect perturbations
of the metric and take $g_{\mu\nu} = \eta_{\mu\nu}$.
We can derive a non-relativistic description in the standard way, by
writing the 4-component spinor $\Psi_{\rm e}$ in a specific Lorentz
frame as
\begin{equation}
\label{transform}
\Psi_{\rm e}
= e^{-i m_{\rm e} c^2 t}
\left( \begin{array}{c}
\Phi  \\
\chi \end{array} \right),
\end{equation}
solving the equation of motion for $\chi$, substituting that
solution back into the action, and discarding relativistic corrections.
The result at leading order is the following action for the electron
field $\Phi$
\begin{equation}
S = \int d^3x dt \, \left[ i \Phi^\dagger {\dot {\Phi}} - \frac{1}{2
    m_{\rm e}} ({\bf \nabla} \Phi)^\dagger ({\bf \nabla} \Phi) + g (
    \Phi^\dagger \Phi)^2 \right],
\label{eq:action12}
\end{equation}
where the coupling constant is $g = - \sqrt{3} m_{\rm e}^2 / ( 48 c m_*^4)$.
The action (\ref{eq:action12})
yields at leading order the following Hamiltonian describing the
interaction of two non-relativistic electrons:
\begin{equation}
{\hat H} = \frac{{\hat {\bf p}}_1^2}{2 m_e} + \frac{{\hat {\bf
      p}}_2^2}{2 m_e} - \frac{e^2}{|{\bf r}_1 - {\bf r}_2|} -
      \frac{\beta m_e^2}{c m_*^4} \delta^3({\bf r}_1 - {\bf r}_2).
\label{eq:ham}
\end{equation}
Here $\beta$ is a dimensionless constant of order unity, ${\bf r}_1$,
${\bf r}_2$, ${\hat {\bf p}}_1$ and ${\hat {\bf p}}_2$ are the
positions and momenta of the two electrons, we have added the
Coulomb interaction term and we have neglected spin effects.
Thus, the leading order correction is a
contact interaction.  The Hamiltonian (\ref{eq:ham}) yields
for the total cross section for backward\footnote{We choose
to focus on this observable since the total forward plus backward
Coulomb scattering cross section diverges.} scattering the conventional result
$\sigma \sim e^4 / E^2$ for $E \ll e^2 m_*^4 c / m_e^3 \sim 10^{-32} \,
{\rm eV}$, where $E$ is
the center of mass energy,
whereas for $E \gg e^2 m_*^4 c/ m_e^3$ we obtain the modified
prediction
\begin{eqnarray}
\sigma &\sim& \frac{m_e^6}{c^2 m_*^8}. \\
\nonumber
\label{eq:prediction}
\end{eqnarray}
Based on this prediction, we claimed in Ref.\ \cite{flanagan} that the
action (\ref{eq:action11}) disagrees with experiments.

However, this claim is not quite correct, because the prediction
(\ref{eq:prediction}) is only valid in an energy regime which is
inaccessible to experiments.  This is so for two
reasons.  First, the higher order corrections indicated by the dots in
Eq.\ (\ref{eq:action11}) give rise to corrections to the action
(\ref{eq:action12}).  For example, there is a term proportional to
$m_{\rm e}^3 (\Phi^\dagger \Phi)^3 / (c^4 m_*^8)$.  These correction terms
become important at energies $E \agt E_{\rm c} = m_* c^2 (m_* / m_{\rm
  e})^{5/3} \sim
10^{-18} \, {\rm eV}$.  At energies above $E_{\rm c}$ an accurate
computation of the tree level cross section would require an infinite
number of terms.
Second, although the theory (\ref{eq:action11}) is a weakly
coupled effective quantum field theory at low energies, there are
indications that above a
certain energy scale it becomes strongly coupled, i.e. loop
corrections become large.  This follows from the fact that the tree
level $s$-wave electron-electron scattering cross section obtained from
the action (\ref{eq:action12}) is of order $\sim m_{\rm
  e}^2 g^2$, which exceeds the unitarity limit $\sim 1/(m_{\rm e} E)$
\cite{weinberg} for energies above
\begin{equation}
E_{\rm sc} \sim \frac{1}{m_{\rm e}^3 g^2} \sim m_* c^2 \left(
\frac{m_*}{m_{\rm e}} \right)^7 \sim 10^{-62} \, {\rm eV}.
\end{equation}
The same conclusion can also be reached by applying dimensional
analysis \cite{polchinski}
to the action (\ref{eq:action12}).  The coupling $g$ has dimensions
$[M]^{-3/2} [E]^{-1/2}$, and so for a process of energy $E$ the
interaction term in the action (\ref{eq:action12}) will give a
contribution to $S$ of order $ g m_e^{3/2} E^{1/2} \sim (E/E_{\rm
  sc})^{1/2}$.  Therefore the
effective description will break down at $E \sim E_{\rm
  sc}$.  Of course, this conclusion might be modified once one
includes in the analysis all of the other fields in the standard model
besides the
electron.  Nevertheless the indication is that the resulting theory
probably becomes strongly coupled at energy scales that are too low to
be accessible to experiments.

Therefore, while one expects large corrections from the extra terms in
the action (\ref{eq:action11}), it is not possible to compute these
corrections reliably at accessible energy scales.  Thus, it is not
possible to definitively rule out the theory.  However, one can make
the following general arguments against the theory (\ref{eq:action4})
as model of the Universe's acceleration.  First,
in this theory one looses the ability to make predictions in atomic
physics and low energy particle physics.  That is, the theory replaces
the standard model of particle physics with a matter field theory
which is strongly coupled at low energies.  Normally, in positing a
modification to gravity one would like to retain the successful
theoretical description of atomic physics provided by the standard
model.  Second, in order to make a quantitative model of the
acceleration of the Universe, one needs to invoke the
approximation that the Universe is homogeneous.  This approximation is
invalid for the theory (\ref{eq:action4}), and so one cannot justify
any Friedmann-Robertson-Walker models \cite{flanagan}.
For these reasons we feel that this theory does not provide a
successful model of the Universe's acceleration.

\medskip

\acknowledgments
We thank Dan Vollick for useful criticisms, and Marc Favata,
Daniel Grumiller, Scott Hughes, Sergei Odintsov, Saul Teukolsky, and
Peng Wang for useful comments on the manuscript.  This research was
supported in part by NSF grant PHY-0140209.

\end{document}